\documentclass[conference]{IEEEtran}
\IEEEoverridecommandlockouts
\usepackage{cite}
\usepackage{amsmath,amssymb,amsfonts}
\usepackage{algorithmic}
\usepackage{graphicx}
\usepackage{textcomp}
\usepackage{xcolor}
\usepackage{diagbox}
\usepackage{url}
\usepackage{multirow}
\usepackage{bm}
\def\BibTeX{{\rm B\kern-.05em{\sc i\kern-.025em b}\kern-.08em
    T\kern-.1667em\lower.7ex\hbox{E}\kern-.125emX}}
\begin{document}

\title{Depthwise Separable Convolutions Versus Recurrent Neural Networks for Monaural Singing Voice Separation
\thanks{K. Drossos and T. Virtanen would like to acknowledge CSC Finland for computational resources.
Stylianos I. Mimilakis is supported in part by the German Research Foundation (AB 675/2-1, MU 2686/11-1).}}

\author{\IEEEauthorblockN{Pyry Pyykk\"{o}nen\IEEEauthorrefmark{1}, Styliannos I. Mimilakis\IEEEauthorrefmark{2}, Konstantinos Drossos\IEEEauthorrefmark{3}, and Tuomas Virtanen\IEEEauthorrefmark{3}} \IEEEauthorblockA{\IEEEauthorrefmark{1}3D Media Research Group, 
Tampere University, Tampere, Finland\\
Email: \{firstname.lastname\}@tuni.fi} \IEEEauthorblockA{\IEEEauthorrefmark{2}Semantic Music Technologies Group, Fraunhofer-IDMT, Ilmenau, Germany\\
Email: mis@idmt.fraunhofer.de}
\IEEEauthorblockA{\IEEEauthorrefmark{3}Audio Research Group, Tampere University, Tampere, Finland\\Email: \{firstname.lastname\}@tuni.fi} }

\maketitle

\begin{abstract}
Recent approaches for music source separation are almost exclusively based on deep neural networks, mostly employing recurrent neural networks (RNNs). Although RNNs are in many cases superior than other types of deep neural networks for sequence processing, they are known to have specific difficulties in training and parallelization, especially for the typically long sequences encountered in music source separation. In this paper we present a use-case of replacing RNNs with depth-wise separable (DWS) convolutions, which are a lightweight and faster variant of the typical convolutions. We focus on singing voice separation, employing an RNN architecture, and we replace the RNNs with DWS convolutions (DWS-CNNs). We conduct an ablation study and examine the effect of the number of channels and layers of DWS-CNNs on the source separation performance, by utilizing the standard metrics of signal-to-artifacts, signal-to-interference, and signal-to-distortion ratio. Our results show that by replacing RNNs with DWS-CNNs yields an improvement of 1.20, 0.06, 0.37 dB, respectively, while using only 20.57\% of the amount of parameters of the RNN architecture. 
\end{abstract}


\section{Introduction}
The task of audio source separation is to extract the underlying audio sources from an observed audio mixture. A particular problem that has attracted great attention in audio and music source separation, is the estimation of the singing voice and accompaniment sources~\cite{rafii18}. To address this problem, a common and successfully employed work flow, consists of computing non-negative signal representations, and employing deep neural networks (DNNs) to estimate the target sources.

Although different methods have been recently proposed for computing and learning signal adaptive/dependent representations for source separation~\cite{adaptive_fe_ss, tzinistwostep, mimilakis2020unsupervised}, the short-time Fourier transform (STFT) remains a popular choice among state-of-the-art (SOTA) approaches in music source separation~\cite{drossos:2018:ijcnn, mim18, spleeter_icassp, stoter19, real_time}. Specifically, by using the STFT, the complex valued representation of the mixture signal is computed. Then, the corresponding magnitude information of the mixture signal is processed by an appropriated method, e.g. DNNs, yielding the magnitude information of the target source. Using the phase information of the mixture, the time-domain signals of the estimated sources are recovered by means of the inverse STFT (ISTFT).

Focusing on the DNNs that estimate the target source in the STFT domain, a certain approach that state-of-the-art methods employ is that of filtering/masking. This approach, enforces DNNs to output filters that are optimized for separating audio and music sources, and has led to good results for both separation quality~\cite{drossos:2018:ijcnn, spleeter_icassp, stoter19} and computational costs~\cite{real_time}. In more details, DNNs are conditioned on the mixture signal magnitude spectrogram and are optimized, in a supervised fashion, to yield a time-varying filter, i.e., a time-frequency mask. The time-frequency mask is applied to the input mixture spectrogram, resulting into a filtered version of the input mixture. The parameters of the DNNs are optimized to minimize the difference between the filtered and the targeted source spectrograms, available in the training dataset. The main benefit of employing such approach versus other approaches is that the DNNs are more efficient in learning the spectrogram structure of the target music source~\cite{mappings2020}.

Typical DNN masking-based approaches for music source separation rely on recurrent neural networks (RNNs) to encode information from the mixture magnitude spectrogram~\cite{drossos:2018:ijcnn, mim18, stoter19}, that is then decoded to obtain the source-dependent mask. However, many previous works have highlighted that
the optimization of the DNNs could be difficult,
due to the involved RNNs, resulting into a very slow, or even sub-optimal learning process. A few reasons to that are improper gradient norms of the RNN parameters during training~\cite{pascanu_13}, and the large number of parameters RNNs require to efficiently process long sequences~\cite{graves_seq_gen}. Although techniques, such as skip-connections~\cite{graves_seq_gen}, bi-directional sequence sampling~\cite{graves_seq_gen}, and regularization schemes~\cite{serdyuk:twinnet} have been proposed to alleviate the above severe issues, CNNs have an increased popularity~\cite{spleeter_icassp, takahashi17, demucs,  drossos:2020:ijcnn}. In contrast to RNNs, CNNs have fewer parameters and can be easily parallelised, resulting into a faster learning process. Furthermore, recent works have shown that depth-wise separable CNNs can even perform better than typical CNNs in a wide range of applications spanning from image recognition~\cite{guo:2018:bmvc} to sound event detection~\cite{drossos:2020:ijcnn} and speech~\cite{tasnet-19} and music source separation~\cite{demucs}. 

Because of the above, in this work we conduct an ablation study and examine the objective performance differences in singing voice separation, by replacing the RNNs with depth-wise separable CNNs. To that aim, we particularly focus on the Masker and Denoiser (MaD) architecture presented in the following works~\cite{mimilakis:2017:mlsp, mim18, drossos:2018:ijcnn}. We do so because MaD architecture incorporates the RNN techniques that have been previously presented in~\cite{mim18} and in~\cite{drossos:2018:ijcnn}, serving a fair, yet competitive baseline for the scope of this work.

The rest of the paper is organized as follows. In Section~\ref{sec:method} we present our proposed method, consisting of the replacement of RNNs with depth-wise separable convolutions at the MaD architecture. In Section~\ref{sec:evaluation} we presented the followed evaluation procedure, and the obtained results are presented in Section~\ref{sec:results}. Section~\ref{sec:conclusions} concludes the paper. 

\section{Proposed method}\label{sec:method}
Our method accepts as an input the magnitude spectrogram $\mathbf{V}\in\mathbb{R}_{\geq0}^{T+L\times F}$ of the musical mixture, consisting of $T+L$ time frames with $F$ frequency bands, and outputs the magnitude spectrogram $\hat{\mathbf{V}}_{j}\in\mathbb{R}_{\geq0}^{T\times F}$ of the $j$-th targeted source, by applying a two-step process. First, our method filters $\mathbf{V}$, producing an initial estimate of the magnitude spectrogram of the $j$-th source, $\hat{\mathbf{V}}'_{j}\in\mathbb{R}_{\geq0}^{T\times N}$, where the extra $L$ vectors of $\mathbf{V}$ are used as temporal context for the initial estimate $\hat{\mathbf{V}}'_{j}$. Then, our method enhances $\hat{\mathbf{V}}'_{j}$, producing the final estimate of the magnitude spectrogram of the $j$-th source, $\hat{\mathbf{V}}_{j}$. 

Our proposed method in based on the MaD system~\cite{mimilakis:2017:mlsp, mim18, drossos:2018:ijcnn}, which takes as an input $\mathbf{V}$ and employs two denoising auto-encoders (DAEs), one for estimating $\hat{\mathbf{V}}'_{j}$, and one for calculating $\hat{\mathbf{V}}_{j}$. The first DAE in MaD is based on RNNs, which are known to be hard to use for parallelized training, and more hard to optimize than CNNs~\cite{drossos:2020:ijcnn, he:2019:jop, lea:2016:eccv}. 

\subsection{MaD system}
MaD consists of two modules; the masker and the denoiser. The masker accepts as an input $\mathbf{V}$ and outputs $\hat{\mathbf{V}}'_{j}$, and it consists of
a trimming operation, $Tr$, 
a bi-directional RNN encoder, $\text{RNN}_{\text{enc}}$, a unidirectional RNN decoder, $\text{RNN}_{\text{dec}}$, and a feed-forward layer, $\text{FNN}_{\text{m}}$. 

The trimming operation, $Tr$, takes as an input $\mathbf{V}$ and reduces the amount of frequency bands from $F$ to $N$, resulting in $\mathbf{V}_{\text{tr}}\in\mathbb{R}_{\geq0}^{T+L\times N}$. This is done in order to reduce the input dimensionality of $\text{RNN}_{\text{enc}}$, consequently reducing the amount of parameters of $\text{RNN}_{\text{enc}}$. Though, the complete $\mathbf{V}$ will be used later on, after the $\text{RNN}_{\text{enc}}$. The bi-directional $\text{RNN}_{\text{enc}}$ consists of a forward RNN, $\overrightarrow{\text{RNN}}_{\text{enc}}$, and a backward RNN, $\overleftarrow{\text{RNN}}_{\text{enc}}$, takes as an input $\mathbf{V}_{\text{tr}}$ and processes it according to
\begin{align}
    \overrightarrow{\mathbf{h}}'^{t'}_{\text{enc}} =& \overrightarrow{\text{RNN}}_{\text{enc}}(\mathbf{v}^{t'}_{\text{tr}}, \overrightarrow{\mathbf{h}}'^{t'-1}_{\text{enc}})\text{ and}\\
    \overleftarrow{\mathbf{h}}'^{t'}_{\text{enc}} =& \overleftarrow{\text{RNN}}_{\text{enc}}(\overleftarrow{\mathbf{v}}^{t'}_{\text{tr}}, \overleftarrow{\mathbf{h}}'^{t'-1}_{\text{enc}})\text{,}
\end{align}
\noindent
where $\overrightarrow{\mathbf{h}}'^{t'}_{\text{enc}},\, \overleftarrow{\mathbf{h}}'^{t'}_{\text{enc}}\in[-1, 1]^{2N}$ are the latent outputs of $\overrightarrow{\text{RNN}}_{\text{enc}}$ and $\overleftarrow{\text{RNN}}_{\text{enc}}$, respectively, at the $t'$-th time frame,
% \begin{equation}
%     \mathbf{h}'^{t'}_{\text{enc}} = \text{RNN}_{\text{enc}}(\mathbf{v}^{t'}_{\text{tr}}, \mathbf{v}^{T+L+1-t'}_{\text{tr}}, \mathbf{h}'^{t'-1}_{\text{enc}})\text{, }
% \end{equation}
% \noindent
% where $\mathbf{h}'^{t'}_{\text{enc}}\in[-1, 1]^{2N}$ is the latent output of $\text{RNN}_{\text{enc}}$ at the $t'$-th time frame, consisting of $\mathbf{h}'^{t'}_{\text{enc}} = [{\overrightarrow{\mathbf{h}}'^{t'}}_{\text{enc}}^{\top}, {\overleftarrow{\mathbf{h}}'^{t'}}_{\text{enc}}^{\top}]^{\top}$, where $\overrightarrow{\mathbf{h}}'_{\text{enc}}\in[-1, 1]^{N}$ and $\overleftarrow{\mathbf{h}}'_{\text{enc}}\in[-1, 1]^{N}$ are the forward and backward (respectively) latent outputs of the forward and backward RNNs of the bi-directional $\text{RNN}_{\text{enc}}$, respectively, 
$t'=1, \ldots,T+L$, $\overrightarrow{\mathbf{h}}'^{0}_{\text{enc}} = \overleftarrow{\mathbf{h}}'^{0}_{\text{enc}} = \{0\}^{N}$, $\overleftarrow{\mathbf{v}}^{t'}_{\text{tr}}$ is the time-flipped (i.e. backwards) version of $\mathbf{V}_{\text{tr}}$, and $\mathbf{H}'_{\text{enc}} = [\mathbf{h}'^{1}_{\text{enc}}, \ldots, \mathbf{h}'^{T+L}_{\text{enc}}]$. Bi-directional $\text{RNN}_{\text{enc}}$ is used to encode the input magnitude spectrogram, using extra information from the $L$ temporal content vectors. The output of the encoder $\mathbf{H}'_{\text{enc}}$, is summed with the input $\mathbf{V}$, using residual connections as
\begin{equation}
    \mathbf{H}_{\text{enc}} = \mathbf{H}'_{\text{enc}} + [\mathbf{V}_{\text{tr}}^{\top}, \overleftarrow{\mathbf{V}}_{\text{tr}}^{\top}]^{\top}\text{,}
\end{equation}
\noindent
where $\overleftarrow{\mathbf{V}}_{\text{tr}}$ is the magnitude spectrogram $\mathbf{V}_{\text{tr}}$ flipped in time (i.e. backwards) and $\mathbf{H}_{\text{enc}}\in\mathbb{R}^{T+L\times2N}$. Finally, the extra $L$ time-frames are dropped from $\mathbf{H}_{\text{enc}}$, so the subsequent decoder will be able to focus on the time frames that correspond to the targeted output, as
\begin{equation}
    \mathbf{H}_{\text{enc-tr}} = [\mathbf{h}^{\lfloor L/2\rfloor}_{\text{enc}}, \ldots, \mathbf{h}^{T + \lfloor L/2\rfloor}_{\text{enc}}]\text{,}
\end{equation}
\noindent
where $\mathbf{h}^{i}_{\text{enc}}\in[-1, 1]^{N}$ is the $i$-th vector of $\mathbf{H}_{\text{enc}}$ and $\lfloor\cdot\rfloor$ is the floor function. $\mathbf{H}_{\text{enc-tr}}$ is used as an input to $\text{RNN}_{\text{dec}}$ of masker, obtaining $\mathbf{H}_{\text{dec}}$ as
\begin{equation}
    \mathbf{h}^{t}_{\text{dec}} = \text{RNN}_{\text{dec}}(\mathbf{h}^{t}_{\text{enc-tr}}, \mathbf{h}^{t-1}_{\text{dec}})\text{,}
\end{equation}
\noindent
where $\mathbf{h}^{t}_{\text{dec}}$ is the latent output of the $\text{RNN}_{\text{dec}}$ at the $t$-th time-frame, $t=1, \ldots, T$, $\mathbf{h}^{0}_{\text{dec}} = \{0\}^{N}$, and $\mathbf{H}_{\text{dec}} = [\mathbf{h}_{\text{dec}}^{1},\ldots,\mathbf{h}_{\text{dec}}^{T}]$. $\mathbf{H}_{\text{dec}}$ is given as an input to a feed-forward linear layer with shared weights through time, followed by a rectified linear unit (ReLU) as
% \begin{equation}
%     \mathbf{H}_{\text{m}} = \text{ReLU}(\text{FNN}_{\text{m}}(\mathbf{H}_{\text{dec}}))\text{,}
% \end{equation}
\begin{equation}
    \mathbf{h}^{t}_{\text{m}} = \text{ReLU}(\text{FNN}_{\text{m}}(\mathbf{h}^{t}_{\text{dec}}))\text{,}
\end{equation}
\noindent
where $\mathbf{h}^{t}_{\text{m}}\in\mathbb{R}_{\geq0}^{F}$ and $\mathbf{H}_{\text{m}}=[\mathbf{h}^{1}_{\text{m}},\ldots,\mathbf{h}^{T}_{\text{m}}]$. Finally, the output of the masker, $\hat{\mathbf{V}}'_{j}$, is calculated as
\begin{equation}\label{eq:masker-output}
    \hat{\mathbf{V}}'_{j} = \mathbf{V}'\odot\mathbf{H}_{\text{m}}\text{,}
\end{equation}
\noindent
where ``$\odot$'' is the Hadamard product and $\mathbf{V}'=[\mathbf{v}^{\lfloor L/2\rfloor},\ldots,\mathbf{v}^{T+\lfloor L/2\rfloor}]$ is a time-trimmed version of the input magnitude spectrogram $\mathbf{V}$ (i.e. before the trimming process $Tr$).

The denoiser, accepts as an input the $\hat{\mathbf{V}}'_{j}$ and outputs $\hat{\mathbf{V}}_{j}$, and it consists of two feed-forward layers with shared weights through time and functioning as an auto-encoder, $\text{FNN}_{\text{d1}}$ and $\text{FNN}_{\text{d2}}$, where each one is followed by a ReLU. Specifically, the first layer, $\text{FNN}_{\text{d1}}$, process the input to the decoder as
\begin{equation}
    \mathbf{h}^{t}_{\text{d1}} = \text{ReLU}(\text{FNN}_{\text{d1}}(\hat{\mathbf{v}}'^{t}_{j}))\text{,}
\end{equation}
\noindent
where $\mathbf{h}^{t}_{\text{d1}}\in\mathbb{R}_{\geq0}^{\lfloor F/2 \rfloor}$ and $\mathbf{H}_{\text{d1}}=[\mathbf{h}^{1}_{\text{d1}},\ldots,\mathbf{h}^{T}_{\text{d1}}]$. Then, the second layer, $\text{FNN}_{\text{d1}}$, process $\mathbf{H}_{\text{d1}}$ as
\begin{equation}
    \mathbf{h}^{t}_{\text{d2}} = \text{ReLU}(\text{FNN}_{\text{d2}}(\mathbf{h}^{t}_{\text{d1}}))\text{,}
\end{equation}
\noindent
where $\mathbf{h}^{t}_{\text{d2}}\in\mathbb{R}_{\geq0}^{F}$ and $\mathbf{H}_{\text{d2}}=[\mathbf{h}^{1}_{\text{d2}},\ldots,\mathbf{h}^{T}_{\text{d2}}]$. The output of the denoiser, $\hat{\mathbf{V}}_{j}$ is calculated as
\begin{equation}
    \hat{\mathbf{V}}_{j} = \hat{\mathbf{V}}'_{j}\odot\mathbf{H}_{\text{d2}}\text{.}
\end{equation}

Finally, the masker and the denoiser are jointly optimized by minimizing
\begin{align}
    \mathcal{L} = &D_{\text{KL}}(\mathbf{V}_{j} \, || \, \hat{\mathbf{V}}'_{j}) + D_{\text{KL}}(\mathbf{V}_{j} \, || \, \hat{\mathbf{V}}_{j})\nonumber\\ &+\lambda_{1}|\text{diag}\{\mathbf{W}_{\text{FNN}_{\text{m}}}\}|_{1}+\lambda_{2}||\mathbf{W}_{\text{FNN\textsubscript{d2}}}||^{2}_{2}\text{,}
\end{align}
\noindent
where $\mathbf{V}_{j}$ is the targeted magnitude spectrogram of the $j$-th source, $D_{\text{KL}}$ is the generalized Kullback-Leibler divergence, $\lambda_{1}=1\times10^{-2}$ and $\lambda_{2}=1\times10^{-4}$ are regularization terms, $|\cdot|_{1}$ is the $\ell_{1}$ vector norm, and $||\cdot||_{2}^{2}$ is the $L_{2}$ matrix norm. $\text{diag}\{\mathbf{W}_{\text{FNN}_{\text{m}}}\}$ is the main diagonal of the weight matrix of the $\text{FNN}_{\text{m}}$ (i.e. the elements $w_{ij}$ of $\mathbf{W}_{\text{FNN}_{\text{m}}}$ with $i=j$). More information about the specific regularizations terms and optimization process, can be found at the original MaD and the MaDTwinNet papers~\cite{mimilakis:2017:mlsp,drossos:2018:ijcnn}.

\subsection{Replacing RNNs} %with depth-wise separable convolutions}
In our proposed method, we replace the bi-directional $\text{RNN}_{\text{enc}}$ and the unidirectional $\text{RNN}_{\text{dec}}$ with two sets of convolutional blocks, $\text{CNN}_{\text{enc}}$ and $\text{CNN}_{\text{dec}}$, respectively. Following recent and SOTA published work~\cite{drossos:2020:ijcnn}, we opt to employ depth-wise separable (DWS) convolutions and not typical convolutions for our CNN blocks. The DWS convolution is a factorized version of the typical convolution, that first applies a spatial-wise convolution, and then a channel-wise convolution. The spatial-wise convolution learns spatial relationships in the input features to the convolution. The channel-wise convolution, learns cross-channel relationships between the channels of the spatial-wise convolution. 

Specifically, each DWS convolution block of our method consists of a CNN (the spatial-wise convolution $\text{CNN}_{\text{d}}$), followed by a leaky ReLU (LReLU), a batch-normalization process, another CNN (the channel-wise convolution $\text{CNN}_{\text{s}}$), and a ReLU, as
\begin{equation}
    \mathbf{H} = \text{ReLU}(\text{CNN}_{\text{s}}(\text{BN}(\text{LReLU}(\text{CNN}_{\text{d}}(\mathbf{X})))))\text{, where}
\end{equation}
\begin{align}
    \mathbf{D}^{c_{\text{i}},x_{h}-K_{\text{h}}, x_{w}-K_{\text{w}}} =& \text{CNN}_{\text{d}}(\mathbf{X}^{c_{\text{i}}}; \mathbf{K}_{\text{d}}^{c_{\text{i}}})\nonumber\\
    =& (\mathbf{K}_{\text{d}}^{c_{\text{i}}} * \mathbf{X}^{c_{\text{i}}})({x_{h}-K_{\text{h}}, x_{w}-K_{\text{w}}})\nonumber\\
    =& \sum\limits_{k_{\text{h}}=1}^{K_{\text{h}}}\sum\limits_{k_{\text{w}}=1}^{K_{\text{w}}}\mathbf{X}^{c_{\text{i}}, x_{h}-k_{\text{h}}, x_{w}-k_{\text{w}}}\mathbf{K}_{\text{d}}^{c_{\text{i}}, k_{\text{h}}, k_{\text{w}}}\text{,}\label{eq:depth-wise}\\
    \mathbf{H}^{c_{\text{o}},\phi_{h}, \phi_{w}} =& \text{CNN}_{\text{s}}(\mathbf{D}^{:,\phi_{h}, \phi_{w}}; \mathbf{K}_{\text{s}}^{c_{\text{o}}})\nonumber\\
    =& \sum\limits_{c_{\text{i}}=1}^{C_{\text{i}}}\mathbf{D}^{c_{\text{i}},\phi_{h}, \phi_{w}}\mathbf{K}_{\text{s}}^{c_{\text{o}}, c_{i}}\text{,}\label{eq:separable}\\
    \text{LReLU}(x) =& \begin{cases}
    x, &\text{ if }x\geq0,\\
    \beta x &\text{ otherwise}
    \end{cases}\text{,}
\end{align}
\noindent
BN is the batch normalization process, $*$ indicates convolution, $\mathbf{D}\in\mathbb{R}^{C'_{i}\times\Phi_{h}\times\Phi_{w}}$ and $\mathbf{K}_{\text{d}}\in\mathbb{R}^{C_{i}\times K_{\text{d}h}\times K_{\text{d}w}}$ are the output and kernel tensors of $\text{CNN}_{\text{d}}$, respectively, $\mathbf{H}\in\mathbb{R}^{C_{o}\times\Phi'_{h}\times\Phi'_{w}}$ and $\mathbf{K}_{\text{s}}\in\mathbb{R}^{C_{o}\times C_{i}}$ are the output tensor and kernel matrix of $\text{CNN}_{\text{s}}$, respectively, and $\beta<1$ is a hyper-parameter. Eq.~\eqref{eq:depth-wise} is used to learn the spatial relationships of the data $\mathbf{X}\in\mathbb{R}^{C_{i}\times X_{h}\times X_{w}}$, and Eq.~\eqref{eq:separable} is used to learn the cross-channel relationships. We employ LReLU according to previous studies using depth-wise separable convolutions~\cite{drossos:2020:ijcnn}. 

Our $\text{CNN}_{\text{enc}}$ consists of one DWS convolution block that is followed by a batch normalization process, a max-pooling operation, and a dropout with $p_{\text{enc}}$ probability, and then of $L_{\text{enc}}$ DWS convolution blocks, with each block followed by a batch normalization process and a dropout with probability $p_{\text{enc}}$ (but no max-pooling operation). The output of each of the $L_{\text{enc}}$ DWS convolution blocks has the same dimensionality as the input. That is, at each of the $L_{\text{enc}}$ DWS convolution blocks, we utilize proper zero padding (i.e. depending on the kernel size) in order not to alter the dimensions of the input. Each block of $\text{CNN}_{\text{enc}}$ gets as an input the output of the previous one, the first gets as an input $\mathbf{V}$, and the last outputs the tensor $\mathbf{H}^{L_{\text{enc}}}\in\mathbb{R}_{\geq0}^{C_{\text{enc}}\times H_{\text{enc}}\times W_{\text{enc}}}$.

$\text{CNN}_{\text{dec}}$ consists of a transposed convolution, followed by two DWS convolution blocks,  batch-normalization and max-pooling processes, a dropout with probability $p_{\text{dec}}$, a CNN, and the $\text{FNN}_{\text{enc}}$. The transposed convolution of $\text{CNN}_{\text{dec}}$ gets as an input the $\mathbf{H}^{L_{\text{enc}}}$, and the $\text{FNN}_{\text{enc}}$ outputs $\mathbf{H}_{\text{m}}$. Finally, the output of the masker of our method is calculated according to Eq.~\eqref{eq:masker-output}. The final audio signal of the output is calculated according to the original MaD paper~\cite{drossos:2018:ijcnn}.

\section{Experimental Procedure}\label{sec:evaluation}
\subsection{Dataset and pre-processing}
We use the development sub-set of Demixing Secret Dataset\footnote{\url{http://www.sisec17.audiolabs-erlangen.de}} (DSD100) for optimizing the parameters of the proposed method, in a supervised fashion. From each multi-track we compute a monaural version of each of the four sources, by averaging the two available channels. Then, we compute the STFT of each monaural signal using a Hamming window of $2049$ samples (46ms) over a step size of $384$ samples (8ms). Each windowed segment is zero-padded to $4096$ samples. After the STFT, we remove the redundant information of the STFT retaining the first $N=2049$ frequency bands, and then compute the absolute values. Then the magnitude spectrogram of the mixture and singing voice are segmented into $B=\lceil M / T \rceil$ sequences, with $T$ being the length of the sequence, and $\lceil\cdot\rceil$ is the ceiling function. Each sequence $b$ is employed as our $\mathbf{V}$ and $\mathbf{V}_{j}$, for the mixture and target source respectively, and overlaps with the preceding one by an empirical factor of $L \times 2$. The overlap factor is used for aggregating context information in the previously described stages of encoding.
\begin{table*}[!ht]
    \centering
    \caption{SDR, SIR, and SAR values, and amount of parameters ($N_{\text{params}}$) for the different amounts of $\text{CNN}_{\text{\textnormal{enc}}}$ blocks ($L_{\text{\textnormal{enc}}}$) and channels of the corresponding kernel ($C_{\text{\textnormal{o}}}$). Values of SDR, SIR, and SAR are presented in \textnormal{dB}. With bold fonts are the values for the combination of $L_{\text{\textnormal{enc}}}$ and $C_{\text{\textnormal{o}}}$, that yields the bigger SDR.% $N_{\text{params}}$ for MaD is 27 195 538.
    }
    \label{tab:chanells-effect}
    \begin{tabular}{c|*{3}c|*{3}c|*{3}c|*{3}c}
    \multicolumn{1}{c}{} & \multicolumn{3}{c}{\textbf{SDR}} & \multicolumn{3}{c}{\textbf{SIR}} & \multicolumn{3}{c}{\textbf{SAR}} & \multicolumn{3}{c}{$\pmb{N_{\text{params}}}$}\\
    \diagbox[width=14em]{Value of $L_{\text{enc}}$}{Value of $C_{\text{o}}$}
      & 64 & 128 & 256 & 64 & 128 & 256 & 64 & 128 & 256 & 64 & 128 & 256\\
    \hline
    5 & 4.47 & 4.84 & 4.91 & 8.11 & 8.07 & 8.59 & 6.71 & 6.74 & 6.98 & 4 783 426 & 4 922 754 & 5 447 170\\
    7 & 4.44 & 4.65 & \textbf{4.94} & 7.83 & 8.23 & \textbf{8.23} & 6.67 & 6.91 & \textbf{7.15} & 4 795 586 & 4 963 458 & \textbf{5 594 114}\\
    9 & 4.46 & 4.57 & 4.88 & 8.14 & 8.06 &8.44 & 6.45 & 6.59 & 6.95 & 4 807 746 & 5 004 162 & 5 741 058\\
    11 & 4.46 & 4.64 & 4.83 & 8.03 & 8.77 & 8.98 & 6.39 & 6.84 & 8.98 & 4 819 906 & 5 044 866 & 5 888 002\\
    13 & 4.59 & 4.80 & 4.72 & 8.17 & 9.02 & 8.41 & 6.40 & 6.56 & 6.97 & 4 832 066 & 5 085 570 & 6 034 946\\
    15 & 4.39 & 4.58 & 4.76 & 8.77 & 8.41 & 8.60 & 6.20 & 6.61 & 7.01 & 4 844 226 & 5 126 274 & 6 181 890\\
    \end{tabular}
\end{table*}

\subsection{Hyperparameters and training of proposed method}
We evaluate our method by conducting an ablation study, employing different amounts of $\text{CNN}_{\text{enc}}$ blocks, $L_{\text{enc}}$, and different number of channels, $C_{\text{o}}$, for our convolutional kernels. Specifically, we employ six different number of $L_{\text{enc}}$, namely 5, 7, 9, 11, 13, and 15, and three different $C_{\text{o}}$, namely 64, 128, and 256. We indicate the amount of $L_{\text{enc}}$ and $C_{\text{o}}$, using a subscript, e.g. $\text{CNN}_{\text{enc-}5,64}$ for the $L_{\text{enc}}=5$ and $C_{\text{o}}=64$ combination. All DWS convolution blocks of $\text{CNN}_{\text{enc}}$ have a square kernel of $K_{\text{d}h}=K_{\text{d}w}=5$. 
% Additionally, to take advantage of the temporal and harmonic context of music, we introduce bigger kernels in the channel-wise convolution of the DWS convolutions. That is, instead $K_{\text{s}h}=K_{\text{s}w}=1$ in Eq.~\eqref{eq:separable}, we use a square kernel of dimensions $K_{\text{s}h}=K_{\text{d}h}=5$ and $K_{\text{s}w}=K_{\text{d}w}=5$. We indicate the usage of the larger kernel in the channel-wise convolution of the DWS as $\text{CNN}'$, e.g. $\text{CNN}'_{\text{enc-}5,64}$ for a DWS block with $L_{\text{enc}}=5$, $C_{\text{o}}=64$, and $K_{\text{s}h}=K_{\text{s}w}=5$. 
At the $\text{CNN}_{\text{dec}}$ we utilize the same $C_{\text{o}}$ amount of channels with the $\text{CNN}_{\text{enc}}$, $K_{\text{d}h}=K_{\text{d}w}=5$, 
% $K_{\text{s}h}=K_{\text{s}w}=1$ for the $\text{CNN}_{\text{enc}}$ and $K_{\text{s}h}=K_{\text{s}w}=3$ for the $\text{CNN}'_{\text{enc}}$ case, 
and a unit stride, and $\beta=1e-2$. The values for $K_{\text{d}h}$ and $K_{\text{d}w}$ are chosen according to previous work that employed DWS convolutions~\cite{drossos:2020:ijcnn} and the value for $\beta$ as the default value for the LReLU in the PyTorch framework. 
%Gradient clipping, TBA

We optimize the parameters of our method following the approach in the original papers of MaD~\cite{mimilakis:2017:mlsp,mim18}, using 100 epochs on the training dataset, with a batch size of 4. We utilized the Adam optimizer for updating the weights of our method, with a learning rate of 1e-4 and for betas we used the values proposed in the original corresponding paper~\cite{adam}. Additionally, we employ a clipping of the gradient $L_{2}$ norm equal to 0.5, similar to the training process of the original MaD system. 
%
% Finally and based on established practises for deep neural networks, we also employ residual connections between the convolutional blocks, as
% \begin{equation}\label{eq:residual}
%     \mathbf{H}_{l_{\text{enc}}} = \text{CNN}_{\text{enc}}(\mathbf{H}_{l_{\text{enc}} - 1} + \mathbf{H}_{l_{\text{enc}} - 2})
% \end{equation}
% \noindent
% where $l_{\text{enc}}$ is the index of the $\text{CNN}_{\text{enc}}$ block. We indicate the usage of residual connections with the suffix ``w/ res''. 
%
The above are implemented using the PyTorch framework, and our code is freely available online\footnote{\url{https://github.com/pppyykknen/mad-twinnet}}.

\subsection{Objective Evaluation}
We compare our method with an established masking based approach to singing voice separation, denoted as the Masker and Denoiser (MaD) architecture and friends, namely the MaD TwinNet~\cite{drossos:2018:ijcnn} and the MaD architecture with the recurrent inference algorithm~\cite{mim18}. The length of the sequences for the MaD and friends is set to $T=60$ timeframes, according to the corresponding papers~\cite{drossos:2018:ijcnn}. 
We focus on those two particular approaches because to the best of our knowledge those approaches are the only ones that \emph{do not} estimate all the other music sources in an attempt to re-fine the estimated singing voice signal~\cite{drossos:2020:ijcnn, mim18, spleeter_icassp, stoter19}. This allows us to clearly examine the potentials of using depth-wise separable convolutional networks for masking based approaches to singing voice separation. 
For assessment, the evaluation subset of DSD100 (50 mixtures and corresponding sources) is used for measuring the objective performance of our method, in terms of signal-to-distortion
(SDR), signal-to-interference (SIR), and signal-to-artifacts (SAR) ratios. The computation of SDR, SIR, and SAR for all the compared methods is performed over overlapping signal segments, following the proposed rules of the official Signal Separation and Evaluation Campaign (SiSEC)~\cite{sisec17}. 

\section{Results and discussion}\label{sec:results}
% \begin{table}[!t]
%     \centering
%     \caption{SDR values for the different amounts of $\text{CNN}'_{\text{\textnormal{enc}}}$ blocks ($L_{\text{\textnormal{enc}}}$) and channels of the corresponding kernel ($C_{\text{\textnormal{o}}}$). Values of SDR are presented in \textnormal{dB}.}
%     \label{tab:chanells-effect}
%     \begin{tabular}{c|cccc}
%      \multicolumn{1}{c}{} &   \multicolumn{3}{c}{Value of $C_{\text{o}}$}  \\
%      Value of $L_{\text{enc}}$ & 64 & 128 & 256 \\
%      \hline
%     5 & 4.73 & 4.64 & 4.93 \\
%     7 & 4.68 & 5.14 & 5.03 \\
%     9 & 4.73 & 4.93 & 4.44 \\
%     11 & 4.73 & 5.04 & 4.34 \\
%     13 & 4.78 & 4.86 & \textbf{5.25} \\
%     15 & 4.73 & 5.15 & 4.77 \\
%     \end{tabular}
% \end{table}
In Table~\ref{tab:chanells-effect} are amount of parameters and the obtained values for the SDR, SIR, and SAR versus the different $L_{\text{enc}}$ and $C_{\text{o}}$. From that table, it can be seen that the increase of $C_{\text{o}}$ has a bigger impact to the obtained SDR, SIR, and SAR, compared to the increase of $L_{\text{enc}}$. That is, the increase at the amount of channels benefits more the obtained SDR, SIR, and SAR, than the increase of the depth of the CNNs. Though, this benefit from $C_{\text{o}}$ could be attributed to the more pronounced effect that $C_{\text{o}}$ has on $N_{
\text{params}}$. From Table~\ref{tab:chanells-effect}, it can be seen that the increase of $C_{\text{o}}$ has more impact on the total amount of parameters $N_{
\text{params}}$, than increasing $L_{\text{enc}}$. Regarding the best performing combination, we focus on the SDR and we consider as best performing the combination of $L_{\text{enc}}=7$ and $C_{\text{o}}=256$.
% , and for both $\text{CNN}_{\text{enc}}$ and $\text{CNN}'_{\text{enc}}$. For both $\text{CNN}_{\text{enc}}$ and $\text{CNN}'_{\text{enc}}$, it can be seen that the increase at the $C_{\text{o}}$ has a bigger impact to the obtained SDR and SAR, compared to the increase of $L_{\text{enc}}$. That is, the increase at the amount of channels benefits more the obtained SDR and SAR than the increase at the depth of the CNNs. Although the same trend is obsed for the SIR, it seems that the highest value is mainly obtained with $C_{\text{o}}=128$. 
%
% Comparing $\text{CNN}_{\text{enc}}$ and $\text{CNN}'_{\text{enc}}$, there is a clear benefit of using bigger kernel of the channel-wise convolution of the DWS convolutions. The difference at the amount of parameters between $\text{CNN}_{\text{enc}}$ and $\text{CNN}'_{\text{enc}}$ is $Q'_{\text{enc}}/Q_{\text{enc}} = K_{\text{s}h}\times K_{\text{s}w}$, where $Q'_{\text{enc}}$ is the amount of parameters of $\text{CNN}'_{\text{enc}}$, and $Q_{\text{enc}}$ of $\text{CNN}_{\text{enc}}$. In our work, the ratio $Q'_{\text{enc}}/Q_{\text{enc}}$ is equal to 25. Thus, there might be the case that the benefit due to the bigger kernel of the channel-wise convolution of the DWS convolutions, is mainly due to greater amount of parameters. Though, more investigation on that woudl provide more conclusive results. The best performance is obtained for the $\text{CNN}'_{\text{enc-}13,256}$, which is 0.68 dB greater than the SDR obtained by MaDTwinNet (i.e. MaD with TwinNet)~\cite{drossos:2018:ijcnn} and 1.63 dB greater than the MaD~\cite{mim18}. 

\begin{table}[!t]
    \centering
    \caption{Comparison of our proposed method with MaD, on DSD dataset. Values of SDR, SIR, and SAR are presented in dB. $N_{\text{params-M}}$ is the amount of parameters for Masker. Results of MaD are taken from the literature.}
    \label{tab:results}
    \begin{tabular}{lcccc}
     \multicolumn{1}{c}{\textbf{Approach}} & \textbf{SDR}  & \textbf{SIR} & \textbf{SAR} & $\pmb{N_{\text{params-M}}}$\\
     \hline
     MaD~\cite{mim18} & 3.62 & 7.06  & 5.88 & \multirow{3}{*}{22 996 113} \\
     MaD-Rec.Inferece~\cite{mim18} & 4.20 & 7.94  & 5.91 & \\  
     MaDTwinNet~\cite{drossos:2018:ijcnn} & 4.57 & 8.17  & 5.95 & \\  
    %  $\text{CNN}'_{\text{enc-}5,64}$ & 4.73 & 9.58 & 5.80\\
    \hline
     $\text{CNN}_{\text{enc-}7,256}$ & 4.94 & 8.23 & 7.15 & 1 394 689\\
    %  $\text{CNN}'_{\text{enc-}5,64}$ w/ res & 4.51 & 9.93 & 5.54 \\
    %  $\text{CNN}'_{\text{enc-}13,256}$ w/ res & \textbf{5.63} & \textbf{10.57} & \textbf{7.50} 4199425
    \end{tabular}
\end{table}

To evaluate the benefit of our proposed method compared to the usage of RNNs, we compare our results with the vanilla, with recurrent inference, and with twin networks variants of the MaD system. In Table~\ref{tab:results} are the SDR, SIR, and SAR values of the best performing combination according to SDR and Table~\ref{tab:chanells-effect} (i.e. $\text{CNN}_{\text{enc-}7,256}$), compared to the values for the same metrics obtained from MaD system. Additionally, since one of the main benefits of DWS convolutions is that they have quite few parameters, we also list in Table~\ref{tab:chanells-effect} the amount of parameters of the Masker. We do not list the parameters for the Denoiser, since the Denoiser is the same in all the listed systems in Table~\ref{tab:chanells-effect}. For reference, the amount of parameters of the Denoiser is 4 199 425. 
% Based on established practises for deep neural networks, we also employ residual connections between the convolutional blocks for our $\text{CNN}'_{\text{enc}}$ blocks, in order to enhance more the performance of our method but without affecting the total amount of parameters. The residual connections are implemented as
% \begin{equation}\label{eq:residual}
%     \mathbf{H}_{l_{\text{enc}}} = \text{CNN}'_{\text{enc}, l_{\text{enc}}}(\mathbf{H}_{l_{\text{enc}} - 1} + \mathbf{H}_{l_{\text{enc}} - 2})\text{,}
% \end{equation}
% \noindent
% where $l_{\text{enc}}$ is the index of the $\text{CNN}'_{\text{enc}}$ block. We indicate the usage of residual connections with the suffix ``w/ res''. The amount of parameters of the $\text{CNN}_{\text{enc-}13,256}$ (i.e. 28.1M) is almost the same as of the MaDTwinNet (i.e. 28M). 

As can be seen from Table~\ref{tab:results}, our proposed method surpassed all variants of the MaD system, while, at the same time, it has only 6\% of the parameters at the Masker (i.e. 94\% reduction) compared to MaD. Specifically, we achieve an increase of 0.37 dB, 0.06 dB, and 1.20 dB for SRD, SIR, and SAR, respectively, when using our method and compared to the MaD system trained with the TwinNet regularization (i.e. MaD TwinNet), which is the best performing variant of MaD. As can be seen, the improvement is mainly attributed on the reduction of artifacts in the separated signal (i.e. increase in the SAR). This indicates that the replacement of the RNNs with DWS convolutions can result in signals that have less distortion from the separation method~\cite{bss_eval}. 

Finally, comparing Tables~\ref{tab:chanells-effect} and~\ref{tab:results}, we can see that with our method, with $L_{\text{enc}}=5$ and $C_{\text{o}}=64$, we can use the 2.54\% of the parameters of the Masker, and still have an increase of 0.76 dB at the SAR, while having a marginal reduction of 0.10 dB and 0.06 dB at SDR and SIR, respectively. Basically, this means that with our method we can significantly reduce the parameters of the Masker by 97.5\%, while still getting some improvement at the reduction of distortion from the separation method (i.e. increase at the SAR). In terms of the amount of total parameters, with the best performing combination of our method, $\text{CNN}_{\text{enc-}7,256}$, we get a reduction of 79.43\% (i.e. we use only 20.57\% of the total MaD parameters), and with the $\text{CNN}_{\text{enc-}5,64}$ we get a reduction of 82.41\% (i.e. we use only 17.59\% of the total MaD parameters).
% By using residual connections, we manage to further increase the obtained values by 1.06 dB, 2.40 dB, and 1.55 dB at SDR, SIR, and SAR, respectively. Finally, a comparison between Tables~\ref{tab:chanells-effect} and~\ref{tab:results} can reveal that our method yields better results than MaD architecture for $C_{\text{o}}\geq128$ and for all $L_{\text{enc}}$, apart from $C_{\text{o}}=128$ and $L_{\text{enc}}=15$. This fact strengthens the indication that the replacement of RNNs with DWS convolutions can result in increasing the performance of the monaural singing voice separation method. 

% We picked the specific corner cases because the one is the case with the smallest numbers for $L_{\text{enc}}$ and $C_{\text{o}}$ (i.e. $\text{CNN}_{\text{enc-}5,64}$) and the other (i.e. $\text{CNN}_{\text{enc-}13,256}$) is the one that provided the best results according to Table~\ref{tab:chanells-effect}. 
% Additionally, the amount of parameters of the $\text{CNN}_{\text{enc-}13,256}$ (i.e. 28.1M) is almost the same as of the MaDTwinNet (i.e. 28M), but the amount of parameters of $\text{CNN}_{\text{enc-}5,64}$ is approximately six times smaller than the amount of parameters of MaDTwinNet (i.e. 5.3M vs 28M).
% Still, both of the corner cases of our approach perform better than MaD with TwinNet and MaD without TwinNet. Additionally, we can see that the residual connections have a considerable improvement over the values of the metrics, yielding an improvement of 1.06, 2.40, and 1.55 dB at SDR, SIR, and SAR, respectively. 

\section{Conclusions}\label{sec:conclusions}
In this work we %presented an approach of replacing the
examined the effect in objective separation performance of replacing RNNs with with depth-wise separable (DWS) CNNs. 
% Additionally, we proposed the usage of bigger kernels in the channel-wise convolution of the depth-wise separable CNNS (i.e. bigger than the $1\times1$), in order to take advantage of the temporal and harmonic patterns in music. 
To assess our proposed approach, we focused on the singing voice separation task and we employed a SOTA performing architecture for monaural singing voice separation that is based on RNNs, we implemented our proposed replacements, and we evaluated the performance of the method with the replacements using an established and freely available dataset for music source separation. We evaluated the performance of the singing voice separation using the widely employed source separation metrics of signal-to-distortion (SDR), signal-to-artifacts (SAR), and signal-to-interference (SIR) ratios. The results show a clear benefit of using our approach, both in the performance and the total amount of parameters needed. Specifically, with our approach we managed to reduce the amount of total parameters by 79.43\%, and achieve an increase of 0.37 dB, 0.06 dB, and 1.20 dB at SDR, SIR, and SAR, compared to the original method with RNNs. For future work, we intend to examine the usage of dilated convolutions, in order to exploit the strong temporal context of music (e.g. melody). Additionally, further investigation could be carried, regarding the benefit or having a bigger kernel at the channel-wise convolution, in the depth-wise separable convolution.

\bibliographystyle{IEEEtran}
\IEEEtriggeratref{24}
\bibliography{refs}

\end{document}